\documentclass[fleqn,11pt,twoside]{article}

\makeatletter

\usepackage{amsthm}
\usepackage{amsmath}
\usepackage{latexsym}
\usepackage{amssymb}
\usepackage{amscd}
\usepackage{epsfig}
\usepackage{graphics}
\usepackage{ifthen}

\newcommand{\ArticleLabel}{Article label}
\newcommand{\evenhead}{Author \ name}
\newcommand{\oddhead}{Article \ name}
\newcommand{\theArticleName}{Article name}
\newcommand{\theAuthorNameForContents}{Author}
\newcommand{\theArticleNameForContents}{Article}

\newcommand{\FirstPageHeading}[2]{
\renewcommand{\ArticleLabel}{#1}
\newpage\thispagestyle{empty}\label{\ArticleLabel-fp}%
\noindent\raisebox{24pt}[0pt][0pt]{\makebox[\textwidth]{\protect\footnotesize \sf 
Proceedings of 4th Workshop ``Group Analysis of Differential Equations \& Integrability'' 
\hfill 2009, \pageref{\ArticleLabel-fp}--\pageref{\ArticleLabel-lp}}}\par}

\newcommand{\LastPageEnding}{\label{\ArticleLabel-lp}\newpage}

\newcommand{\AuthorNameForContents}[1]{\renewcommand{\theAuthorNameForContents}{#1}}
\newcommand{\ArticleNameForContents}[1]{\renewcommand{\theArticleNameForContents}{#1}}
\newcommand{\ArticleName}[1]{\renewcommand{\theArticleName}{#1}\vspace{-2mm}\par\noindent {\LARGE\bf  #1\par}}
\newcommand{\Author}[1]{\vspace{5mm}\par\noindent {\it #1} \par\vspace{2mm}\par}
\newcommand{\Address}[1]{\vspace{2mm}\par\noindent {\it #1} \par}
\newcommand{\Email}[1]{\ifthenelse{\equal{#1}{}}{}{\par\noindent {\rm E-mail: }{\it  #1} \par}}
\newcommand{\EmailD}[1]{\ifthenelse{\equal{#1}{}}{}{\par\noindent {$\phantom{\dag}$~\rm E-mail: }{\it  #1} \par}}
\newcommand{\Abstract}[1]{\vspace{6mm}\par\noindent\hspace*{8mm}%
\parbox{120mm}{\small #1}\par\vspace{6mm}
\addtocontents{top}{\small\hangindent=10mm\hangafter=1\protect{\it \theAuthorNameForContents}, 
\protect{\rm \theArticleNameForContents}\dotfill\pageref{\ArticleLabel-fp}
\par\vspace{1mm}\par}\par
\addtocontents{tor}{\small\hangindent=10mm\hangafter=1\protect{\it \theAuthorNameForContents}, 
\protect{\rm \theArticleNameForContents}\dotfill\pageref{\ArticleLabel-fp}
\par\vspace{1mm}\par}\par
}
\newcommand{\tableofpapers}{\@starttoc{top}}
\newcommand{\tableofpapersA}{\@starttoc{tor}}

\newcommand{\ShortArticleName}[1]{\renewcommand{\oddhead}{#1}}
\newcommand{\AuthorNameForHeading}[1]{\renewcommand{\evenhead}{#1}}

\renewcommand{\@evenhead}{
\hspace*{-3pt}\raisebox{-15pt}[\headheight][0pt]{\vbox{\hbox to \textwidth 
{\thepage \hfil \evenhead}\vskip4pt \hrule}}}
\renewcommand{\@oddhead}{
\hspace*{-3pt}\raisebox{-15pt}[\headheight][0pt]{\vbox{\hbox to \textwidth 
{\oddhead \hfil \thepage}\vskip4pt\hrule}}}
\renewcommand{\@evenfoot}{}
\renewcommand{\@oddfoot}{}

\long\def\@makecaption#1#2{%
  \vskip\abovecaptionskip
  \sbox\@tempboxa{\small \textbf{#1.}\ \ #2}%
  \ifdim \wd\@tempboxa >\hsize
    {\small \textbf{#1.}\ \ #2}\par
  \else
    \global \@minipagefalse
    \hb@xt@\hsize{\hfil\box\@tempboxa\hfil}%
  \fi
  \vskip\belowcaptionskip}

\def\numberwithin#1#2{\@ifundefined{c@#1}{\@nocounterr{#1}}{%
  \@ifundefined{c@#2}{\@nocnterr{#2}}{%
  \@addtoreset{#1}{#2}%
  \toks@\@xp\@xp\@xp{\csname the#1\endcsname}%
  \@xp\xdef\csname the#1\endcsname
    {\@xp\@nx\csname the#2\endcsname
     .\the\toks@}}}}

{\theoremstyle{definition}

}

\renewenvironment{thebibliography}[1]     
{\medskip
 \list{\@biblabel{\@arabic\c@enumiv}}%
           {\parsep=-0.2ex%
            \settowidth\labelwidth{\@biblabel{#1}}%
            \leftmargin\labelwidth
            \advance\leftmargin\labelsep
            \@openbib@code
            \usecounter{enumiv}%
            \let\p@enumiv\@empty
            \renewcommand\theenumiv{\@arabic\c@enumiv}}%
      \sloppy      \clubpenalty4000
      \@clubpenalty \clubpenalty
      \widowpenalty4000%
      \sfcode`\.\@m}
     {\def\@noitemerr
       {\@latex@warning{Empty `thebibliography' environment}}%
      \endlist}

\makeatother

\newcommand{\p}{\partial}

\newcommand{\todo}[1][\null]{\ensuremath{\clubsuit}}
\newcommand{\const}{\mathop{\rm const}\nolimits}

\begin{document}

\FirstPageHeading{BIHLO}

\ShortArticleName{Symmetries in atmospheric sciences} 

\ArticleName{Symmetries in atmospheric sciences}

\Author{Alexander BIHLO~$^\dag$}
\AuthorNameForHeading{A. BIHLO}
\AuthorNameForContents{BIHLO A.}
\ArticleNameForContents{Symmetries in atmospheric sciences}

\Address{$^\dag$~Faculty of Mathematics, University of Vienna, Nordbergstra{\ss}e 15,\\
$\phantom{^\dag}$ A-1090 Vienna, Austria}
\EmailD{alexander.bihlo@univie.ac.at}

\Abstract{Selected applications of symmetry methods in the atmospheric sciences are reviewed briefly. In particular, focus is put on the utilisation of the classical Lie symmetry approach to derive classes of exact solutions from atmospheric models. This is illustrated with the barotropic vorticity equation. Moreover, the possibility for construction of partially-invariant solutions is discussed for this model. A further point is a discussion of using symmetries for relating different classes of differential equations. This is illustrated with the spherical and the potential vorticity equation. Finally, discrete symmetries are used to derive the minimal finite-mode version of the vorticity equation first discussed by E. Lorenz (1960) in a sound mathematical fashion.
}

\section{Introduction}

Dynamic meteorology is concerned with the mathematical theory of atmospheric motion. It lies the corner stone for daily weather prediction as it prepares the grounds for numerical computer models, without which reliable forecasts are hardly imaginable. However, since the advent of capable supercomputers and the accompanying shift towards methods for exhausting their capacities, analytical investigations of the governing equations have somewhat taken a back seat.

However, as evident, numerical models need benchmark tests to check for their reliability and hence exact solutions of the underlying mathematical models are still of great value. Testing whether a forecast model is able to reproduce a known exact solution of the original equation may serve as a first consistency check. It follows that techniques for obtaining such solutions in a systematic way are of rather pretty importance. For this purpose, the classical Lie symmetry methods are well-suited.

Lots of dynamical models in use in the atmospheric sciences are adapted forms of the Navier-Stokes or the ideal Euler equations, taking into account both the rotation of the earth and the anisotropy of the atmosphere (the region of interest for weather prediction extends about 10 km in the vertical but several thousands of kilometers in the horizontal direction). For large-scale dynamics (i.e., horizontal length scale about some 1000 km) or for sake of conceptual simplicity, it is possible to restrict oneself to two-dimensional models. The most relevant example of such a model in dynamic meteorology is the barotropic vorticity equation. It is derived from the incompressible Euler equations in a rotating reference frame by using the stream function (or vorticity) as dynamic variable.

In this contribution, we discuss symmetries and group-invariant solutions of the barotropic vorticity equation. Moreover, the construction of partially-invariant solutions is shown. These issues are addressed in section \ref{Bihlo:section:Lie}. Subsequently, in section \ref{Bihlo:section:Rotation} the usage of symmetries for finding related differential equations is demonstrated using the potential and the spherical vorticity equation. It is shown that in both cases the rotational term in the equation can be canceled using suitable point transformations. In section \ref{Bihlo:section:FiniteMode} discrete symmetries of the vorticity equation are used to systematically re-derive the Lorenz (1960) model. This contribution ends with a short summary and discussion of future research plans, which can be found in the final section \ref{Bihlo:section:Conclusion}.

\section{Group-invariant solutions and barotropic vorticity\\ equation}\label{Bihlo:section:Lie}

It is worth considering the Lie symmetry problem of the barotropic vorticity equation since to the best of our knowledge, this equation has not been investigated thoroughly in the light of symmetries before. Either only the symmetries and some exact solutions were computed without reference to classification of inequivalent subgroups \cite{Bihlo:Huang&Lou2004,Bihlo:Ibragimov1995} or the classification itself was not done in most complete fashion \cite{Bihlo:Bihlo2007}.

Using the stream function--vorticity notation, the barotropic vorticity equation in Cartesian coordinates reads:
\begin{equation}\label{Bihlo:equation:vorticitybeta}
  \zeta_t+\psi_x\zeta_y-\psi_y\zeta_x+\beta\psi_x=0, \qquad \zeta = \psi_{xx} + \psi_{yy},
\end{equation}
where $\zeta$ stands for the vorticity, $\psi$ is the stream function and $\beta=\const$ is a parameter controlling the north-south variation of the earth angular rotation. It can be expressed as the north-south change of the vertical Coriolis parameter $f=2\Omega\sin\varphi$ via $\beta = \mathrm{d} f/\mathrm{d} y$, where $\Omega$ is the absolute value of the earth angular rotation vector and $\varphi$ denotes the latitude.

Equation~\eqref{Bihlo:equation:vorticitybeta} admits the infinite-dimensional maximal Lie invariance algebra~$\mathcal{B}^\infty_\beta$ generated by the operators 
\begin{align*}
    &\mathcal{D} = t\p_{t} - x\p_{x} - y\p_{y} - 3\psi\p_{\psi}, \qquad \p_t, \qquad \p_{y},\\
    &\mathcal{X}(f)= f(t)\p_{x} - f'(t)y\p_{\psi}, \qquad \mathcal{Z}(g)= g(t)\p_{\psi},
\end{align*}
where $h$, $f$ and $g$ are arbitrary real-valued time-dependent functions. For a systematic group-invariant reduction of the vorticity equation by means of using subalgebras of $\mathcal{B}^\infty_\beta$, it is necessary to compute the corresponding optimal system of inequivalent subalgebras first \cite{Bihlo:Olver1986}. For one-dimensional subalgebras it reads~\cite{Bihlo:Bihlo2007,Bihlo:Bihlo&Popovych2009}
\begin{align*}
  \langle \mathcal{D} \rangle,\qquad \langle \p_t+c\p_y\rangle,\qquad \langle\p_y + \mathcal{X}(f) \rangle,\qquad \langle \mathcal{X}(f)+\mathcal{Z}(g) \rangle,
\end{align*}
where $c=\{0,\pm 1\}$. For two-dimensional subalgebras the optimal system is~\cite{Bihlo:Bihlo&Popovych2009}
\begin{align*}
\begin{split}
    &\left\langle \mathcal{D},\ \p_t \right\rangle,\;\  \left\langle \mathcal{D},\ \p_y + a\mathcal{X}(1) \right\rangle,\;\  \left\langle \mathcal{D},\ \mathcal{X}(|t|^a) + c\mathcal{Z}(|t|^{a-2})\right\rangle,\;\  \left\langle \mathcal{D},\ \mathcal{Z}(|t|^{a-2})\right\rangle, \\
    &\left\langle \p_t + b\p_y,\ \mathcal{X}(e^{a t})+ \mathcal{Z}((a bt+c)e^{a t})\right\rangle,\quad \left\langle \p_t + b\p_y,\ \mathcal{Z}((a bt+c)e^{a t})\right\rangle, \\
    &\left\langle \p_y + \mathcal{X}(f^1),\ \mathcal{X}(1) + \mathcal{Z}(g^2)\right\rangle, \quad \left\langle \p_y + \mathcal{X}(f^1),\ \mathcal{Z}(g^2)\right\rangle, \\
    & \left\langle \mathcal{X}(f^1)+\mathcal{Z}(g^1),\ \mathcal{X}(f^2)+\mathcal{Z}(g^2)\right\rangle,
\end{split}
\end{align*}
with $a,b,c$ being arbitrary constants and $f^i$ and $g^i$, $i=1,2$, being arbitrary real-valued functions of time. In the fifth and sixth subalgebra, additionally the condition $abc=0$ has to hold. In the last subalgebra the pair of functions $(f^1,g^1)$ and $(f^2,g^2)$ has to be linear independent. Using the above two-dimensional subalgebras allows to reduce (\ref{Bihlo:equation:vorticitybeta}) to ODEs. Moreover, it is straightforward to show that the third and the fourth case of one-dimensional subalgebras lead to completely integrable PDEs. Hence, since all but the first two-dimensional subalgebra are extensions of special forms of the third or fourth one-dimensional subalgebras. That is, only reduction by means of the first two-dimensional subalgebra may give an essential new result.

Based on the above classification, it is possible to derive classes of inequivalent group-invariant solutions \cite{Bihlo:Bihlo&Popovych2009}. As a notable example, we investigate reduction by means of $\langle\p_y + \mathcal{X}(f) \rangle$. The invariant functions of this subalgebra are
\[
      p = x-fy, \qquad q = t, \qquad v = \psi + \frac{1}{2}f'y^2,
\]
which allows one to reduce (\ref{Bihlo:equation:vorticitybeta}) to the classical Klein-Gordon equation 
$
    \tilde v_{\tilde p\tilde q} + \beta \tilde v = 0,
$
where
\[
\tilde q = \int \frac{\mathrm{d} q}{1+f^2},\qquad \tilde p = p,\qquad \tilde v = v - \frac{f''}{\beta}p + \frac{h(q)}{\beta} + \frac{((1+f^2)f'')'}{\beta^2}.
\]
For the meteorological application, the case $f=\const$ and making a harmonic ansatz for $\tilde v$ is the most relevant: The corresponding solution represents the classical Rossby wave, which to a large extend governs the weather regimes in the mid-latitudes.

\medskip

\noindent Considering (\ref{Bihlo:equation:vorticitybeta}) as a system of two equations using $\psi$ and $\zeta$ as dependent variables, it is possible to compute partially-invariant solutions \cite{Bihlo:Ovsiannikov1982}. This should be illustrated using the subalgebra $\langle \mathcal{X}(1), \mathcal{Z}(g) \rangle$. Due to the presence of the basis element $\mathcal{Z}(g)$, no ansatz for $\psi$ can be chosen. However, it is still possible to make an ansatz for $\zeta$. Because of the joint invariance under $\mathcal{X}(1)$ and $\mathcal{Z}(g)$ we have $\zeta = \zeta(t,y)$ and thus (\ref{Bihlo:equation:vorticitybeta}) transforms to
\[
    \zeta_t + \psi_x(\zeta_y + \beta) = 0, \qquad \zeta = \psi_{xx} + \psi_{yy}.
\]
Introducing the absolute vorticity $\eta=\zeta+\beta y$ we have to distinguish two cases. For $\eta_y=0$ the solution of the vorticiy equation is
\[
    \psi = \Psi(t,x,y) - \frac{1}{6}\beta y^3 + \frac{1}{2}\eta(t) y^2,
\]
where $\Psi$ satisfies the Laplace equation $\Psi_{xx}+\Psi_{yy}=0$. For $\eta_y\ne 0$ the solution reads
\[
    \psi = \frac{1}{(g^1)^2}F(\omega) - \frac{1}{6}\beta y^3 - \frac{g^1_ty+g_t^0}{g^1}x + f^1y + f^0,
\]
where $\omega = g^1y + g^0$ and $g^1,g^0,f^1,f^0$ are all real-valued functions of time.

\section{Symmetries and the ineffective earth rotation}\label{Bihlo:section:Rotation}

\subsection{The spherical vorticity equation}

Although the vorticity equation in Cartesian coordinates allows one to study some prominent features of large-scale geophysical fluid dynamics, it does not take into account for the earth's sphericity. For this purpose, it is necessary to study vorticity dynamics in a rotating spherical coordinate system. The corresponding equation then is \cite{Bihlo:Platzman1960}:
\begin{equation}\label{Bihlo:equation:vorticitysphere}
    \zeta_t + \frac{1}{a^2}\left(\psi_\lambda\zeta_\mu - \psi_\mu\zeta_\lambda\right) + \frac{2\Omega}{a^2}\psi_\lambda = 0, 
\end{equation}
where $\psi$ is the (spherical) stream function and $\zeta$ the (spherical) vorticity,
\[
\zeta = \frac{1}{a^2}\left[\frac{1}{1-\mu^2}\psi_{\lambda\lambda} + \left((1-\mu^2)\psi_\mu\right)_\mu\right].
\] 
\looseness=-1
Rather than using $\lambda$ (longitude) and $\varphi$ (latitude) as spatial variables, it is advantageous to use $\lambda$ and $\mu = \sin\varphi$. The mean radius of the earth is denoted by~$a$.

The maximal Lie invariance algebra $\mathfrak g_\Omega$ of~(\ref{Bihlo:equation:vorticitysphere}) is generated by the basis operators 
\begin{align*}
     &\mathcal{D} = t\p_{t} - (\psi-\Omega\mu)\p_{\psi} - \Omega t\p_{\lambda}, \qquad \p_{t}, \qquad \mathcal{Z}(g) = g(t)\p_{\psi},\qquad \mathcal{J}_{1} = \p_{\lambda}, \\
     &\mathcal{J}_{2} = \mu\frac{\sin(\lambda+\Omega t)}{\sqrt{1-\mu^2}}\p_{\lambda}+\frac{\cos(\lambda+\Omega t)}{\sqrt{1-\mu^2}} \left((1-\mu^2)\p_{\mu} + \Omega\p_{\psi}\right), \\ 
     &\mathcal{J}_{3} =\mu \frac{\cos(\lambda+\Omega t)}{\sqrt{1-\mu^2}}\p_{\lambda}-\frac{\sin(\lambda+\Omega t)}{\sqrt{1-\mu^2}} \left((1-\mu^2)\p_{\mu} + \Omega\p_{\psi}\right),    
\end{align*}
where $g$ runs through the set of smooth functions of~$t$. 
It is straightforward to map the algebra $\mathfrak g_\Omega$ with $\Omega\ne0$ to the algebra $\mathfrak g_0$ by means of the transformation
\[
    \tilde t = t, \quad \tilde\mu = \mu, \quad \tilde\lambda = \lambda + \Omega t, \quad \tilde\psi = \psi - \Omega\mu.
\]
Moreover, this transformation also maps the equation (\ref{Bihlo:equation:vorticitysphere}) with $\Omega\ne0$ to the equation of the same form with $\Omega=0$. That is, it is possible to disregard the rotational term by setting $\Omega=0$ for all practical calculations and finally obtain the corresponding results for the case $\Omega\ne 0$ by applying the above transformation. We note in passing that this transformation was already used by Platzman \cite{Bihlo:Platzman1960} to transform the vorticity equation to a zero angular momentum coordinate system.

For optimal systems of one- and two-dimensional subalgebras of $\mathfrak g_0$ and the computation of group-invariant solutions of (\ref{Bihlo:equation:vorticitysphere}) in a fashion similar to the previous section, see \cite{Bihlo:Bihlo&Popovych2009}.

\subsection{The potential vorticity equation}

An extension of the classical barotropic vorticity equation is given through the potential vorticity equation. For the flat topography, the equation reads
\begin{equation}\label{vort}
   \zeta_t - F\psi_t + \psi_x\zeta_y-\psi_y\zeta_x + \beta\psi_x = 0, \qquad \zeta = \psi_{xx}+\psi_{yy},
\end{equation}
where $F$ is the ratio of the characteristic length scale to the Rossby radius of deformation \cite{Bihlo:Pedlosky1987}. In the Lagrangian view, the above equation can be understood as individual conservation of the potential vorticity $q=\zeta+f-F\psi$. In this light, the extension to the barotropic vorticity equation, which in turn states the individual conservation of absolute vorticity $\eta=\zeta+f$, becomes most obvious: While barotropic dynamics takes place solely in two dimensions, the additional term $-F\psi$ accounts for a variation of the fluid height in the vertical. In this respect, the potential vorticity equation is especially suited for shallow water theory.

Determining the symmetries for the case $\beta=0$ and $\beta\ne0$ shows that both algebras and hence also the two corresponding equations are mapped to each other by applying the transformation \cite{Bihlo:Bihlo&Popovych2008b}
\[
  \tilde t = t, \qquad \tilde x = x + \frac{\beta}{F}t, \qquad \tilde y = y, \qquad \tilde \psi = \psi - \frac{\beta}{F}y.
\]
Again, this shows that there are models in the atmospheric sciences, in which the rotational terms are apparently not of prime importance.

\section{Symmetries and finite-mode models}\label{Bihlo:section:FiniteMode}

There is a long history in dynamic meteorology to convert the governing nonlinear PDEs to systems of coupled ODEs by means of series expansions, followed by a reasonable truncation of the series in use. Among one of the first models that was analyzed in this way again was the barotropic vorticity equation in Cartesian coordinates. This was done by Lorenz \cite{Bihlo:Lorenz1960} in an ad-hoc fashion: He first of all expanded the vorticity in a a double Fourier series on the torus
\[
    \zeta = \sum_{\mathbf{m}}C_{\mathbf{m}}\exp(i\mathbf{\hat m} \cdot\mathbf{x}), 
\]
where  $\mathbf{x}=x\mathbf i+y\mathbf j$, $\mathbf{m}=m_1\mathbf i+m_2\mathbf j$, $\mathbf{\hat m}=m_1k\mathbf i+m_2l\mathbf j$,
$k=\const$, $l=\const$, $\mathbf i=(1,0,0)^{\rm T}$, $\mathbf j=(0,1,0)^{\rm T}$, $m_1$ and $m_2$ run through the integers and the coefficient $C_{00}$ vanishes. Afterwards he substituted this expansion in (\ref{Bihlo:equation:vorticitybeta}) for $\beta=0$ yielding the spectral form of the vorticity equation:
\begin{equation} \label{Bihlo:equation:specvort}
    \frac{\mathrm{d}C_\mathbf{m}}{\mathrm{d}t} = -\sum_{\mathbf{m}'\ne\mathbf{0}}\frac{c_{\mathbf{m}'}C_{\mathbf{m}-\mathbf{m}'}}{\mathbf{\hat m}'{}^2}\left(\mathbf{k}\cdot[\mathbf{\hat m'}\times\mathbf{\hat m}]\right),
\end{equation}
where $\mathbf{k}=(0,0,1)^{\rm T}$. Finally, he restricted the indices of the Fourier coefficients $C_\mathbf{m}=1/2(A_\mathbf{m}-iB_\mathbf{m})$ to only run through the indices $\{-1,0,1\}$, leading to an eight-component initial model. He then noted that if the imaginary parts $B_\mathbf{m}$ of the coefficients at hand vanish initially, they will remain zero for all times. Moreover, if the real coefficient $A_{11}=-A_{1,-1}$ initially, this equality is also preserved under the finite-mode dynamics. These observations allow to reduce (\ref{Bihlo:equation:specvort}) to the following three-component model
\begin{align}\label{Bihlo:equation:Lorenz1960}
\begin{split}
    \frac{\mathrm{d}A}{\mathrm{d}t} &= -\left(\frac{1}{k^2}-\frac{1}{k^2+l^2}\right)klFG,  \\
    \frac{\mathrm{d}F}{\mathrm{d}t} &= \left(\frac{1}{l^2}-\frac{1}{k^2+l^2}\right)klAG,  \\
    \frac{\mathrm{d}G}{\mathrm{d}t} &= -\frac{1}{2}\left(\frac{1}{l^2}-\frac{1}{k^2}\right)klAF.
\end{split}
\end{align}
where, $A := A_{01}, F := A_{10}, G := A_{1,-1}$.

Now it is instructive to see, whether it is possible to give a more rigorous justification of the two observations by Lorenz \cite{Bihlo:Bihlo&Popovych2008a}. The key for this investigation is to start with the generators of admitted mirror symmetries of (\ref{Bihlo:equation:vorticitybeta}) for $\beta=0$
\begin{align*} 
    e_1\colon& \quad (x,y,t,\psi) \mapsto (x,-y,t,-\psi),\\
    e_2\colon& \quad (x,y,t,\psi) \mapsto (-x,y,t,-\psi),\\
    e_3\colon& \quad (x,y,t,\psi) \mapsto (x,y,-t,-\psi).
\end{align*}
and induce them in the space of Fourier coefficients by means of series expansion. The induced symmetries are
\begin{align*}
  e_1\colon& \quad C_{m_1m_2} \mapsto -C_{m_1,-m_2}, \\
  e_2\colon& \quad C_{m_1m_2} \mapsto -C_{-m_1m_2}, \\
  e_3\colon& \quad C_{m_1m_2} \mapsto -C_{m_1m_2}, \quad t \mapsto -t.
\end{align*}Moreover, the translations with the values $\pi/k$ and  $\pi/l$ in directions of $x$ and $y$, respectively, induce proper transformations of the Fourier coefficients:
\[
  p\colon \quad C_{m_1m_2} \mapsto (-1)^{m_1}C_{m_1m_2}, \qquad q\colon \quad C_{m_1m_2} \mapsto (-1)^{m_2}C_{m_1m_2}.
\]
The transformations $e_1$, $e_2$, $p$ and $q$ act on dependent variable and thus may be useful for reducing the number of Fourier coefficients at hand. Therefore, the finite dimensional symmetry group that is relevant for us is generated by these four elements and has the structure $G\simeq\mathbb{Z}_2\oplus\mathbb{Z}_2\oplus\mathbb{Z}_2\oplus\mathbb{Z}_2$. Selecting various subgroups of $G$ allows to reduce the eight-component initial model to fife-, four- and three-component submodels, respectively. The three-component submodel by Lorenz is derived upon using the subgroup $S=\{1, p  q e_1, p q e_2, e_1e_2\}$. The transformation $e_1e_2$ accounts for his first observation, i.e. $B_{m_1m_2}=0$, while the transformation $pqe_1$ yields the identification $A_{11}=-A_{1,-1}$, which justifies the second observation. In this way (\ref{Bihlo:equation:Lorenz1960}) is derived merely using symmetry techniques.

This example shows that it is possible to derive a consistent finite-mode model by means of a sound mathematical method. Having such a method at ones disposal is especially useful since up to now there are only few criteria for the selection of modes in finite-mode models.

\section{Conclusion and future plans}\label{Bihlo:section:Conclusion}

In this paper, we have reviewed some possible applications of well-established symmetry methods in the atmospheric sciences. The technique of constructing group-invariant solutions allows to systematically re-derive the well-known Rossby wave solution of the barotropic vorticity equation. Moreover, relating differential equations by mapping their Lie algebras to each other enables one to cancel the terms due to the earth's angular rotation in both the potential and spherical vorticity equation. Form the physical viewpoint, this result is somewhat astonishing since these rotational terms are in fact to a great extend responsible for the overall complexity in formation of weather pattern.

In future works, the classical method of Lie reduction should be applied to more sophisticated models of geophysical fluid dynamics, including the quasi-geostrophic model \cite{Bihlo:Holton2004} and some classical convection model \cite{Bihlo:Chandrasekhar1981}. Moreover, the usage of symmetries for a systematic determination of conservation laws in atmospheric science should be examined. This again has potential application in providing consistency checks for numerical integration schemes.

\subsection*{Acknowledgements}

The continuous help and encouragement of Prof.\ Roman Popovych is kindly acknowledged. The author is a recipient of a DOC-fellowship of the Austrian Academy of Science.

\LastPageEnding


\begin{thebibliography}{99}

\footnotesize

\bibitem{Bihlo:Bihlo2007}
Bihlo~A., {\it Solving the vorticity equation with Lie groups},
Wiener Meteorologische Schriften {\bf 6}, 88pp, 2007.

\bibitem{Bihlo:Bihlo&Popovych2008a}
Bihlo~A. and Popovych~R.O., {Symmetry justification of Lorenz maximum simplification},
arXiv:0805.4061v2, 8pp, 2008.

\bibitem{Bihlo:Bihlo&Popovych2008b}
Bihlo~A. and Popovych~R.O., {Symmetry analysis of barotropic potential vorticity equation},
arXiv:0811.3008v1, 6pp, 2008.

\bibitem{Bihlo:Bihlo&Popovych2009}
Bihlo~A. and Popovych~R.O., {Lie symmetries and exact solutions of barotropic vorticity equation},
arXiv:0902.4099v1, 11pp, 2009.

\bibitem{Bihlo:Chandrasekhar1981}
Chandrasekhar~S., {\it Hydrodynamic and hydromagnetic stability},
Dover, New York, 1981.

\bibitem{Bihlo:Holton2004}
Holton~J.R., {\it An introduction to dynamic meteorology},
Elsevier, Amsterdam, 2004.

\bibitem{Bihlo:Huang&Lou2004}
Huang~F. and Lou~S.Y., {Analytical investigation of Rossby waves in atmospheric dynamics},
{\em Phys. Lett. A.} {\bf 320}, 428--437, 2004.

\bibitem{Bihlo:Ibragimov1995}
Ibragimov~N.H.~ed., {\it CRC handbook of Lie group analysis of differential equations},
CRC Press, Boca Raton, 1995

\bibitem{Bihlo:Lorenz1960} 
Lorenz~E., {Maximum simplification of the dynamic equations}, 
{\it Tellus} {\bf 12}, 243--254, 1960.

\bibitem{Bihlo:Olver1986}
Olver~P., {\it Applications of Lie groups to differential equations},
Springer-Verlag, New York, 1986.

\bibitem{Bihlo:Ovsiannikov1982}
Ovsiannikov~L.V., {\it Group analysis of differential equations},
Acad. Press, San Diego, 1982.

\bibitem{Bihlo:Pedlosky1987}
Pedlosky~J., {\it Geophysical fluid dynamics},
Springer-Verlag, New York, 1987.

\bibitem{Bihlo:Platzman1960}
Platzman~G.W., {The spectral form of the vorticity equation},
{\em J. Meteor.} {\bf 17}, 635--644, 1960.

\end{thebibliography}
\end{document}